\documentclass{PoS}

\usepackage{url}

\title{Using bottomonium production as a tomographic probe of the quark-gluon plasma}

\ShortTitle{Using bottomonium production as a tomographic probe of the quark-gluon plasma}

\author{Michael Strickland\\
        Kent State University\\
        E-mail: \email{mstrick6@kent.edu}}

\abstract{
The suppression of bottomonia in ultrarelativistic heavy-ion collisions is a smoking gun for the production of a strongly interacting final state in ultrarelativistic heavy-ion collisions.  Furthermore, these final state interactions are consistent with the production of a hot hydrodynamically expanding quark-gluon plasma with initial temperatures on the order of 600-700 MeV at LHC collision energies.  Models which incorporate plasma screening and bound-state breakup based on high-temperature quantum chromodynamics are in good agreement with the centrality, transverse-momentum, and rapidity dependence of the suppression seen in the highest energy LHC Pb-Pb collisions.  The models of heavy-quark bound state breakup/formation are coupled to the soft dynamics of the quark-gluon plasma using 3+1d relativistic anisotropic hydrodynamics.  At later times, excited state feed-down is taken into account using independently constrained excited state feed down fractions.  Comparisons with LHC experimental data are consistent with primordial suppression of all bottomonium states, with states having the lowest binding energies being the ones which suffer the most suppression.  In this proceedings contribution, I will review recent work to (a) include the effects of regeneration on bottomonium production and (b) to assess the effects of a rapidly changing in-medium potential on bottomonium production in the quark-gluon plasma.
}

\FullConference{13th International Workshop in High pT Physics in the RHIC and LHC Era (High-pT2019)\\
		19-22 March 2019\\
		Knoxville, Tennessee, USA}

\begin{document}

\section{Introduction}

The ongoing ultrarelativistic heavy-ion collision (URHIC) experiments at Brookhaven National Laboratory's (BNL) Relativistic Heavy Ion Collider (RHIC) and the European Organization for Nuclear Research's (CERN) Large Hadron Collider (LHC) probe the properties of the quark-gluon plasma (QGP).  In the QGP colored partons are not confined to live inside hadrons, but can instead propagate over large distances compared to hadronic scales.  One expects to generate a QGP in high temperature and/or high net-baryon density environments.  In the highest energy collisions at RHIC and the LHC one probes the high temperature and low net-baryon density region of the quantum chromodynamics (QCD) phase diagram.  In this region of the QCD phase diagram, lattice gauge theory calculations have found that there is a smooth (crossover) transition from the hadronic phase to the QGP phase at a pseudo-critical temperature of approximately $T_{\rm pc} \simeq 155$ MeV \cite{Bazavov:2013txa,Borsanyi:2016bzg}.  Once the temperature of the QGP exceeds this temperature, light-quark bounds states such as the pion quickly disassociate, however, lattice and perturbative QCD calculations find that heavy-quark bound states such as the $J/\Psi$ and the $\Upsilon$ can survive to much higher temperatures before breaking up.  In practice, one finds that the $J/\Psi$ and $\Upsilon(1s)$ states survive to temperatures of approximately 300-400 MeV and 600-700 MeV, respectively \cite{Andronic:2015wma,Mocsy:2013syh,Brambilla:2010cs}.  

In the seminal papers of Karsch, Matsui, and Satz (KMS) they made the first predictions that heavy quarkonia would ``melt'' in the QGP~\cite{Matsui:1986dk,Karsch:1987pv}.  In addition, they demonstrated that the precise way in which heavy quarkonium bound states melted could provide insight into properties of the QGP such as its initial temperature.
The model used by KMS was based on a non-relativistic potential which is justified in the large quark mass limit.  This can be made more formal using effective field theory methods to integrate out different energy/momentum scales resulting in potential-based non-relativistic QCD (pNRQCD) \cite{PhysRevD.21.203,Lucha:1991vn,Brambilla:2004jw,Brambilla:2010xn}.  Based on direct high-temperature quantum field theory calculations and effective field theory calculations, it is now known that the in-medium heavy-quark potential is complex-valued, with the real part mapping to the typical Debye-screen potential and the imaginary part being related to the in-medium breakup rate of heavy-quark bound states \cite{Laine:2006ns,Dumitru:2007hy,Brambilla:2008cx,Burnier:2009yu,Dumitru:2009fy,Dumitru:2009ni,Margotta:2011ta,Guo:2018vwy}.   The imaginary part of the potential makes the quantum evolution non-unitary (non-Hermitian Hamiltonian), which can be understood in the context of open quantum systems in which there is a heavy-quark bound state coupled to a thermal heat bath \cite{Akamatsu:2011se,Akamatsu:2012vt,Akamatsu:2014qsa,Katz:2015qja,Brambilla:2016wgg,Kajimoto:2017rel,Brambilla:2017zei,Blaizot:2017ypk,Blaizot:2018oev,Yao:2018nmy}.

Models which incorporate plasma screening and bound-state breakup based on high-temperature quantum chromodynamics are in quite good agreement with the centrality, $p_T$, and $y$ dependence of the suppression seen at the highest LHC collision energies.  The models of heavy-quark breakup/reformation are coupled to the soft-sector of quark-gluon plasma using 3+1d anisotropic hydrodynamics \cite{Florkowski:2010cf,Martinez:2010sc,Bazow:2013ifa,Alqahtani:2017jwl,Alqahtani:2017tnq,Alqahtani:2017mhy,Almaalol:2018gjh} to describe the evolution of the system's temperature and viscous corrections~\cite{Dumitru:2009ni,Margotta:2011ta,Strickland:2011mw,Strickland:2011aa,Strickland:2012cq,Krouppa:2015yoa,Krouppa:2016jcl,Krouppa:2017jlg}.  In parallel, there have been applications of the T-matrix approach to heavy quarkonium suppression and regeneration which also have good agreement with the bottomonium suppression data \cite{Grandchamp:2005yw,Rapp:2008tf,Emerick:2011xu,Du:2017qkv,Du:2019tjf}.  
In both approaches, excited state feed-down is taken into account after QGP kinetic freezeout.  Comparisons of both model's predictions with LHC experimental data are consistent with primordial suppression of all bottomonia states including the ground state, with the states having the lowest binding energies being the ones which suffer the most suppression.  In the proceedings, I will review some recent advances in the theory and phenomenology of heavy quarkonium suppression.

\section{Methodology}

In order to model the dissipative dynamics of the QGP we use 3+1d anisotropic hydrodynamics \cite{Alqahtani:2017mhy}.  Anisotropic hydrodynamics is a dissipative relativistic hydrodynamics framework which resums and infinite number of terms in the fluids inverse Reynolds number, allowing it to describe the early-time non-equilibrium dynamics of the QGP into which bottomonium are formed more accurately.  From these simulations once sees that the largest local rest frame pressure anisotropy comes from the difference of the longitudinal and transverse pressures ($P_L/P_T < 1$).  To take this longitudinal to transverse momentum-space anisotropy into account we introduce an anisotropy parameter $\xi$ and a local-rest-frame one-particle distribution function of the form $f(p,x) = f_{\rm eq\!\!}\left(\sqrt{p_T^2 + [1+\xi(x)]p_z^2}\Big/\Lambda(x)\right)$, where $-1 \leq \xi(x) < \infty$ is the local spheroidal momentum-space anisotropy parameter and $\Lambda(x)$ is the local transverse temperature \cite{Romatschke:2003ms,Romatschke:2004jh}.  This form takes into account the difference between the transverse and longitudinal pressures, which is the most important viscous correction generated in heavy-ion collisions.

For the primary model discussed herein we will use the internal-energy-based potential specified originally in Ref.~\cite{Strickland:2011aa}.  In the model, the real part of the potential is obtained from the internal energy of the heavy quark/anti-quark system
\begin{eqnarray}
\Re[V] &=&  -\frac{a}{r} \left(1+\mu \, r\right) e^{-\mu \, r }
+ \frac{2\sigma}{\mu}\left[1-e^{-\mu \, r }\right]
- \sigma \,r\, e^{-\mu \, r } -  \frac{0.8\,\sigma}{m_b^2 r} 
\, , \label{eq:real_pot_model_B}
\end{eqnarray}
with $m_b = 4.7$ GeV, $a=0.385$, $\sigma = 0.223\;{\rm GeV}^2$ \cite{Petreczky:2010yn}.  The last term is a temperature- and spin-independent finite-quark-mass correction taken from Ref.~\cite{Bali:1997am}.  In this expression, $\mu = {\cal G}(\xi,\theta) m_D$ \cite{Dumitru:2007hy,Dumitru:2009ni,Strickland:2011aa} is the anisotropic Debye mass, where ${\cal G}$ is a function which depends on the degree of plasma momentum-space anisotropy $\xi$, the angle of the line connecting the quark-antiquark pair with respect to the beamline direction $\theta$, and the isotropic Debye mass $m_D = 1.4 \sqrt{1+N_f/6} \, g_s T$ \cite{Kaczmarek:2004gv}.\footnote{In the limit $\xi \rightarrow 0$, one has ${\cal G} = 1$ and the real part of the potential above reduces to the internal energy derived from the original Karsch-Mehr-Satz potential~\cite{Karsch:1987pv}.}. The imaginary part of the potential $\Im[V]$ is taken from a leading-order perturbation theory calculation in the small-$\xi$ limit~\cite{Laine:2006ns,Dumitru:2009fy,Burnier:2009yu,Nopoush:2017zbu}
\begin{eqnarray} 
\Im[V] &=& - \alpha_s C_F T \, \Big\{ \phi(r/m_D)  - \xi \left[ \psi_1(r/m_D,\theta)+\psi_2(r/m_D, \theta)\right] \Big\} \, ,
\label{eq:impot}
\end{eqnarray}
where $\phi$, $\psi_1$, and $\psi_2$ are special functions which can be expressed in terms of the Meijer G-function.  We close by emphasizing that, since the local temperature varies in time, the potential above is inherently time dependent.  We will now discuss two approaches to this problem.

\subsection{The adiabatic limit}

In the first approach, one assumes that the time scale for variation of the potential is long compared to the timescale of the internal quantum dynamics.  This gives the {\em adiabatic limit}.  In this limit one can calculate the instantaneous breakup rate by solving a time-independent Schr\"odinger equation to obtain the real and imaginary parts of the bound-state energy as a function of the local temperature $T$ and momentum-anisotropy $\xi$ \cite{Margotta:2011ta,Strickland:2011aa} and then combine these results with the 3+1d hydrodynamical simulation results for $T({\bf x},t)$ and $\xi({\bf x},t)$.  To obtain the suppression (survival probability ignoring regeneration) one uses
\begin{eqnarray}
&&R_{\rm AA}(p_T,y,{\bf x}_\perp,b) = e^{-\zeta(p_T,y,{\bf x}_\perp,b)} \, , \nonumber \\
&&\zeta \equiv \Theta(\tau_f-\tau_{\rm form}) \int_{{\rm max}(\tau_{\rm form},\tau_0)}^{\tau_f} 
\! d\tau \; \Gamma(\tau,{\bf x}_\perp,\varsigma=y) \, , \;\;\;\;\;
\label{eq:raa}
\end{eqnarray}
where $b$ is the heavy-ion impact parameter, $\tau_{\rm form} = \gamma \tau_{\rm form}^0 = E_T \tau_{\rm form}^0/M$ where $M$ is the mass of the state, and $\tau_{\rm form}^0$ is the formation time of the state at rest.  The function $\Gamma$ is the breakup rate which is determined separately as function of the temperature and anisotropy parameter from solution to the 3d Schr\"odinger equation~\cite{Strickland:2011aa}.  For the rest frame formation times, we assume that they are proportional to the inverse vacuum binding energy~\cite{Karsch:1987uk}.  For the $\Upsilon(1s)$, $\Upsilon(2s)$, $\Upsilon(3s)$, $\chi_{b1}$, and $\chi_{b2}$ states, we use $\tau_{\rm form}^0$ = 0.2, 0.4, 0.6, 0.4, and 0.6 fm/c, respectively.  Below we will refer to this model as KSUa.

\subsubsection*{Results}

\begin{figure}[t!]
\centerline{
\includegraphics[width=0.485\linewidth]{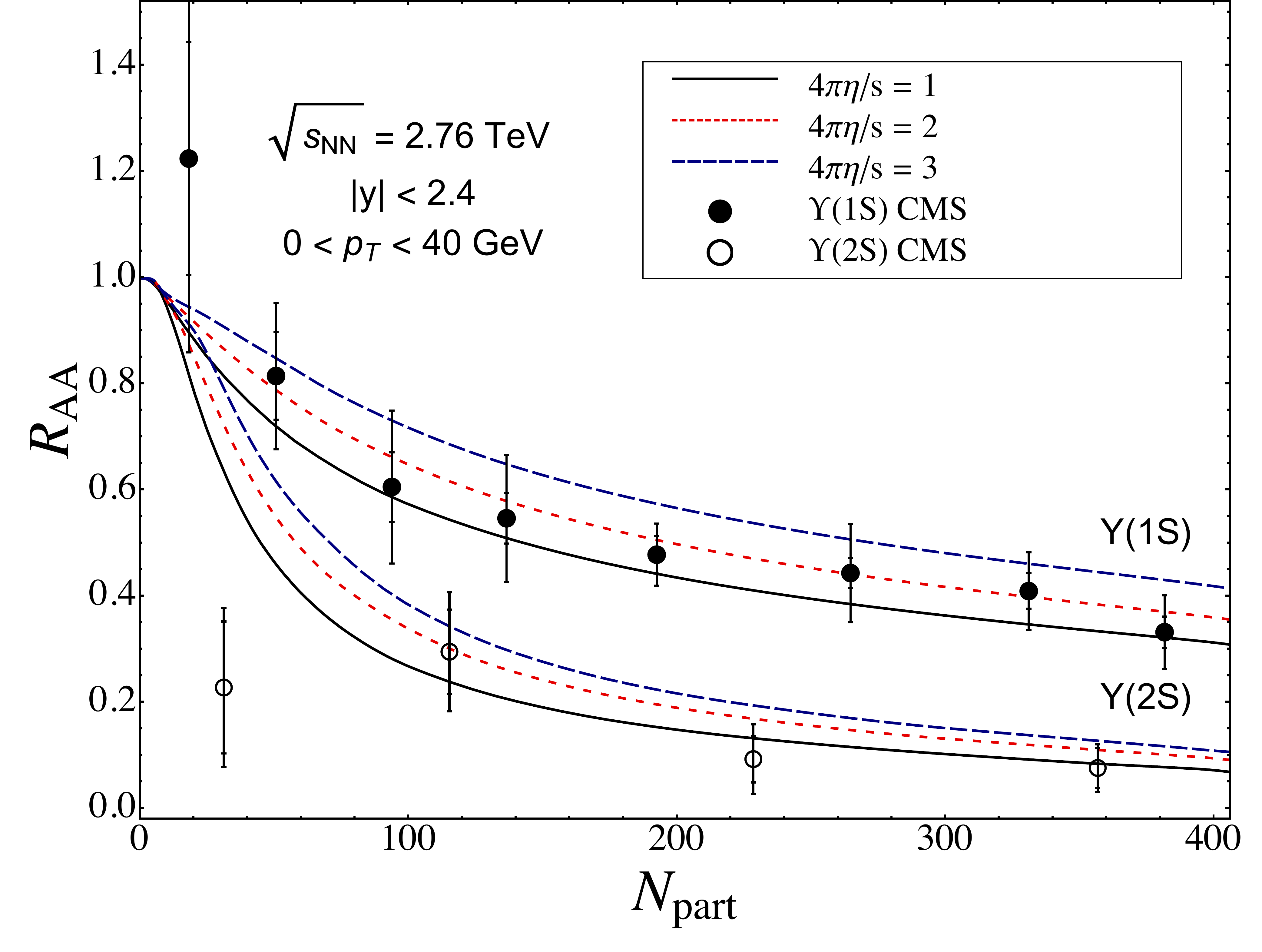}\hspace{2mm}
\includegraphics[width=0.485\linewidth]{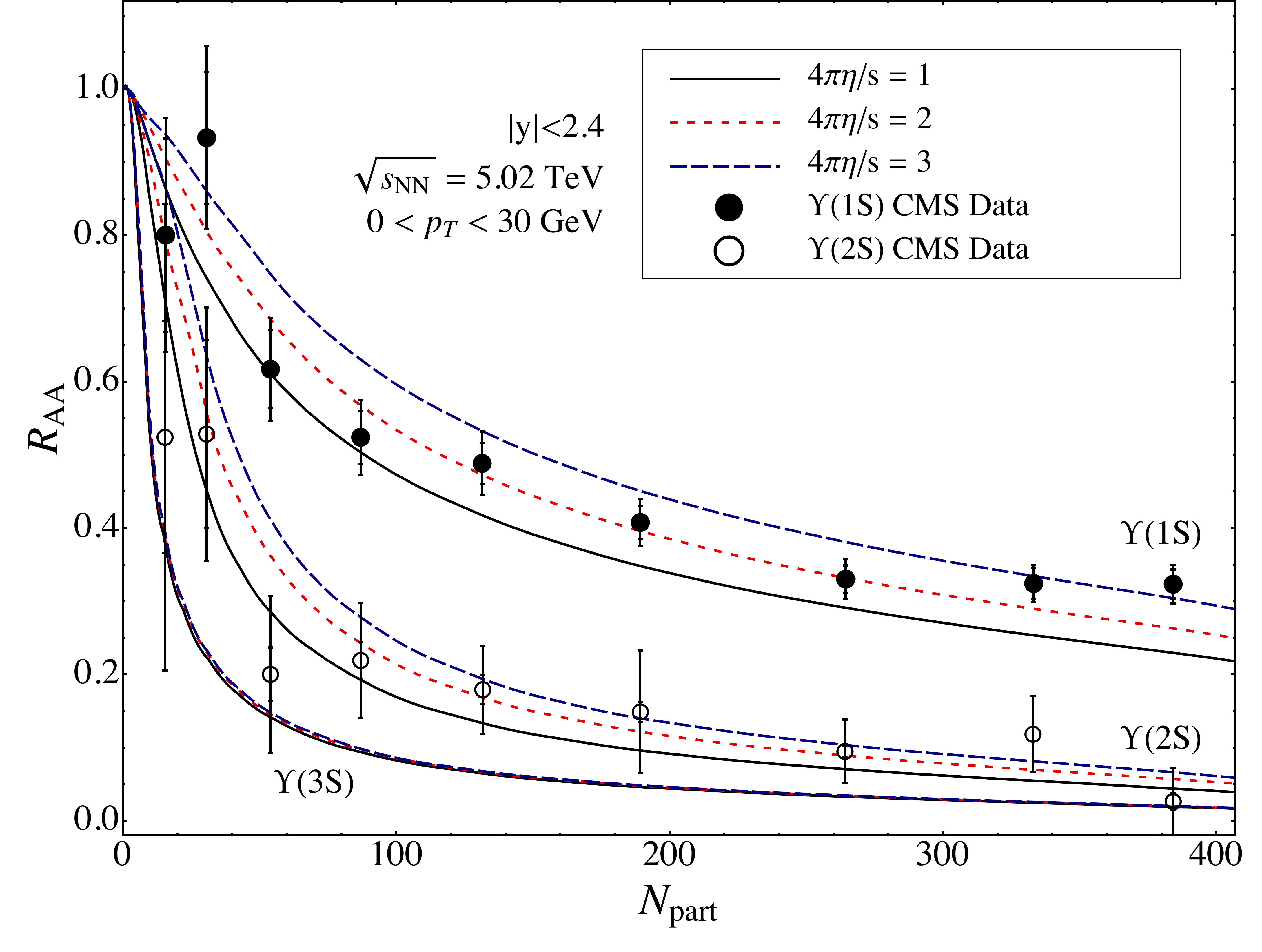}
}
\caption{$R_{AA}$ versus $N_{\rm part}$ for $\Upsilon(1S)$ and $\Upsilon(2S)$.  Left panel shows $\sqrt{s_{NN}} = 2.76$ TeV with CMS data from \cite{Khachatryan:2016xxp} and the right panel shows $\sqrt{s_{NN}} = 5.02$ TeV with preliminary CMS data from \cite{CMS5TeV}.  Lines correspond to three different values of the shear viscosity to entropy density ratio.}
\label{fig:1}
\end{figure}

\begin{figure}[t!]
\centerline{
\includegraphics[width=0.485\linewidth]{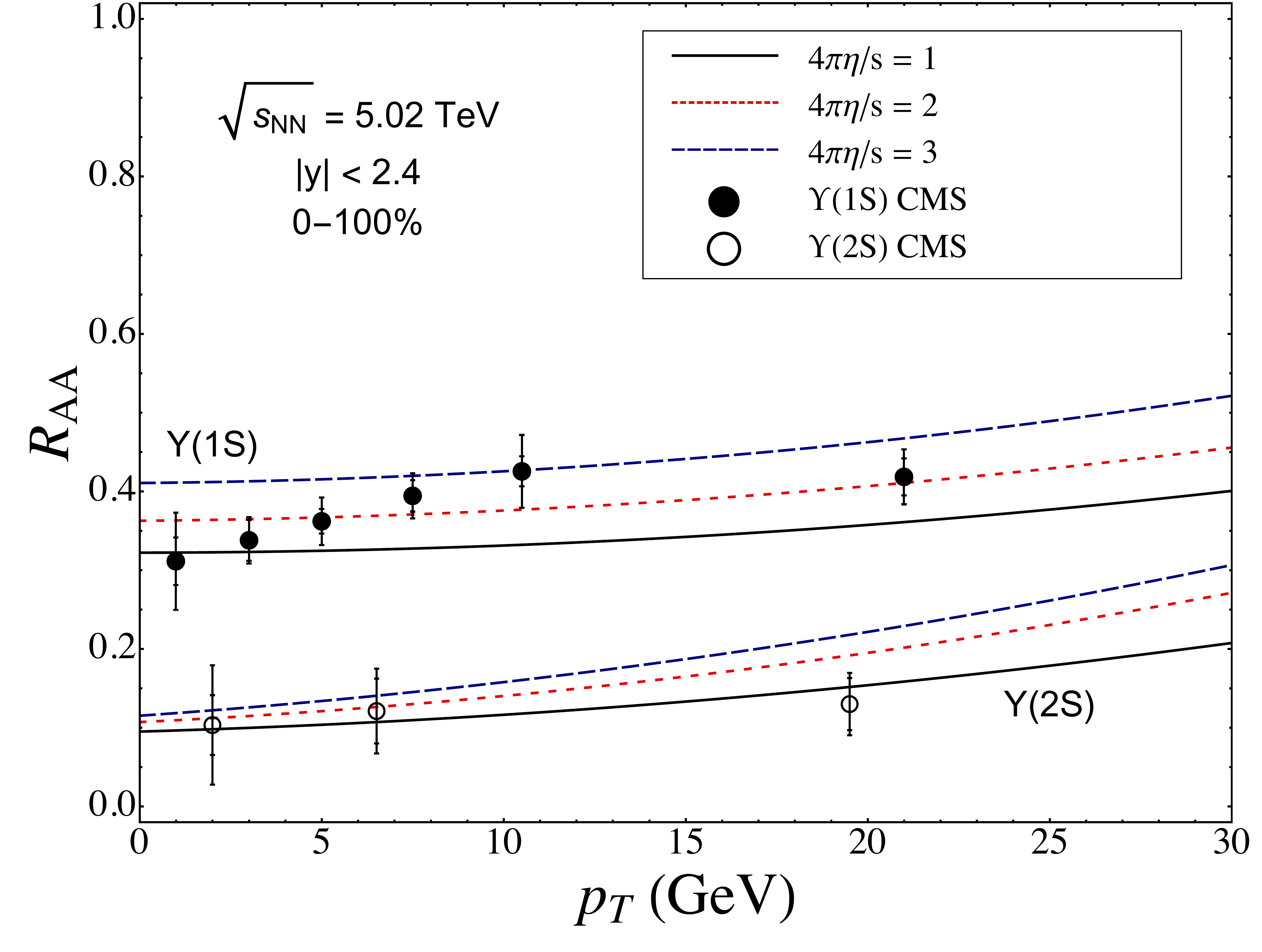}\hspace{2mm}
\includegraphics[width=0.485\linewidth]{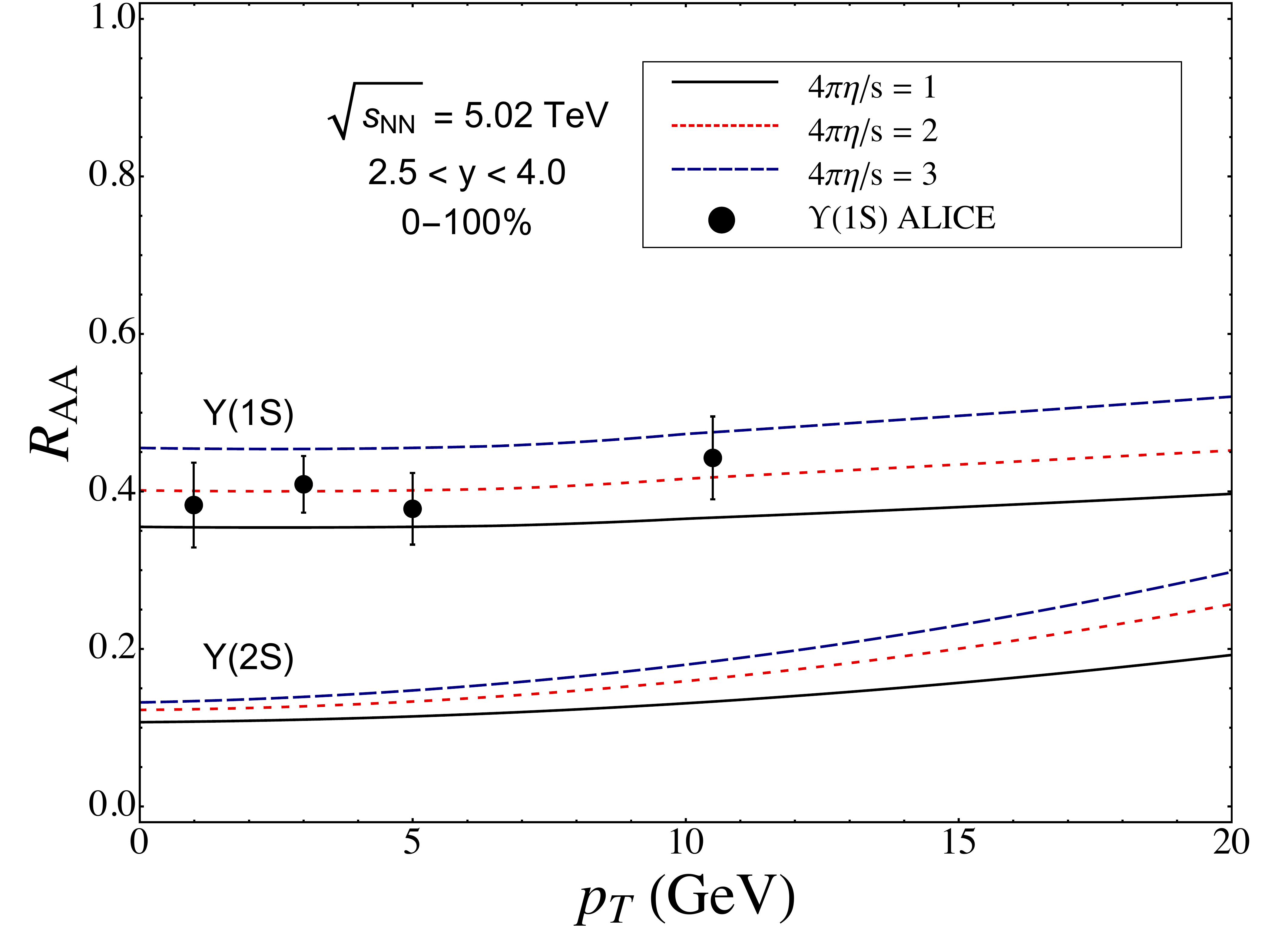}
}
\caption{$R_{AA}$ versus $p_T$ for $\Upsilon(1S)$ and $\Upsilon(2S)$ at $\sqrt{s_{NN}} = 5.02$ TeV.  Left panel shows preliminary CMS data from \cite{CMS5TeV} and the right panel shows preliminary ALICE data from \cite{ALICE5TeV}.  Lines are same as Fig.~\ref{fig:1}.}
\label{fig:2}
\end{figure}

In Fig.~\ref{fig:1} we plot the $N_{\rm part}$ dependence of bottomonium suppression.  On the left we show the theoretical predictions for $\Upsilon(1s)$ and $\Upsilon(2s)$ suppression for 2.76 TeV Pb-Pb collisions.  The experimental data is from the CMS collaboration  \cite{Khachatryan:2016xxp} and the long-dashed blue, dashed red, and solid black lines correspond to different assumed values of the shear viscosity to entropy density ratio, $\eta/s$.  For each value of $\eta/s$, the initial central temperature was adjusted in order keep the total charged hadron multiplicity fixed.  On the right we show the same at 5.02 TeV along with predictions for the $\Upsilon(3s)$ suppression.  The experimental data for this figure was taken from Ref.~\cite{CMS5TeV}.  As this figure demonstrates, the KSUa model works well to describe the centrality dependence of the observed $R_{AA}$ of the $\Upsilon(1s)$ and $\Upsilon(2s)$ well at both collision energies.

In Fig.~\ref{fig:2} we show the transverse-momentum dependence of the $\Upsilon(1s)$ and $\Upsilon(2s)$ $R_{AA}$.  The left plot shows a comparison with CMS data \cite{CMS5TeV} collected in the rapidity interval $|y| < 2.4$ and the right plot shows a comparison with ALICE data \cite{ALICE5TeV} collected in the rapidity interval $2.5 < y < 4.0$.  Once again, we see that the KSUa model describes the data quite well in both rapidity windows.  In the model, the decrease in the suppression with increasing transverse momentum comes solely from the time dilation of the bound state formation time.

\begin{figure}[t!]
\centerline{
\includegraphics[width=0.485\linewidth]{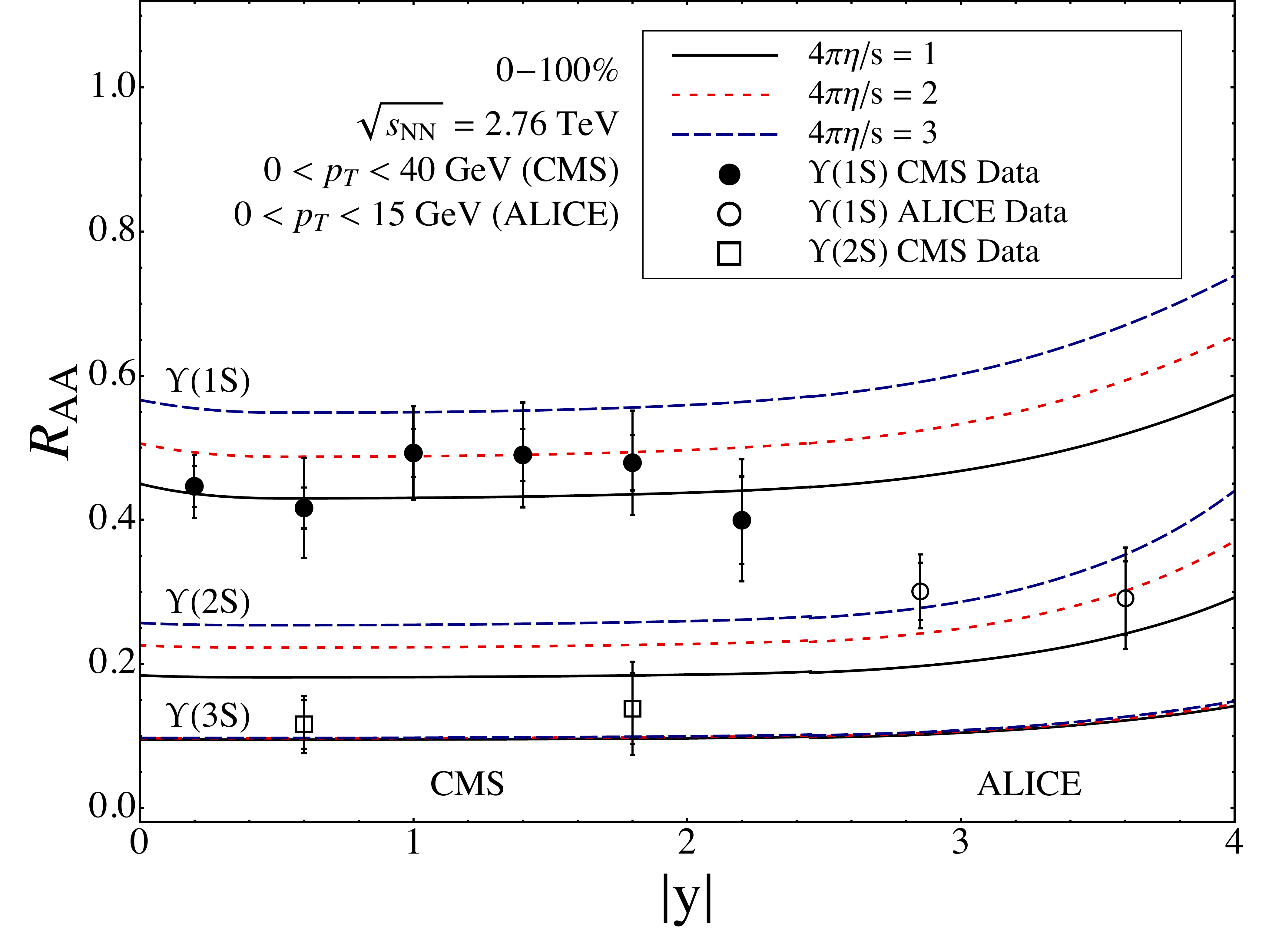}\hspace{2mm}
\includegraphics[width=0.485\linewidth]{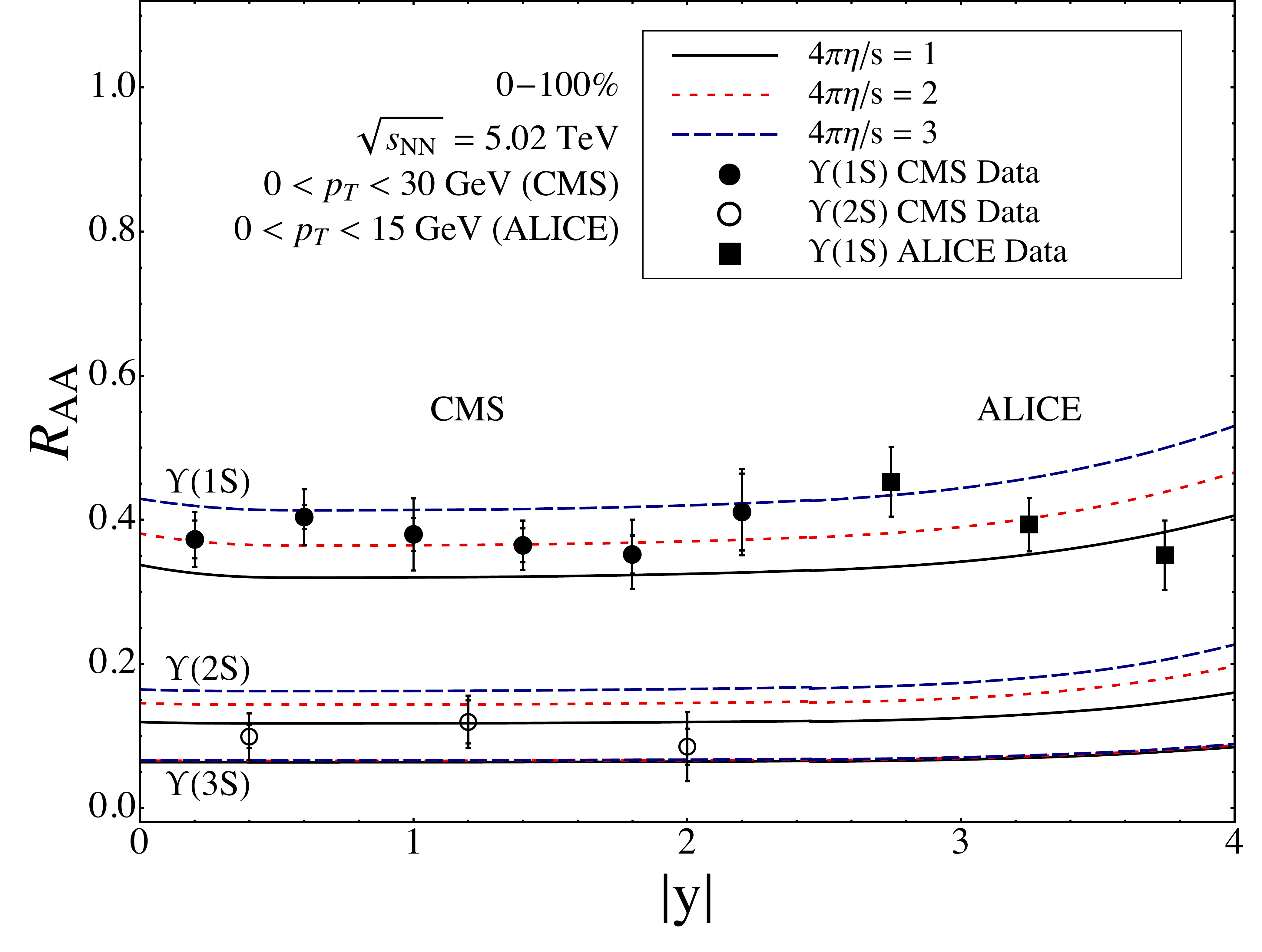}
}
\caption{(Left) $R_{AA}$ versus $y$ for $\Upsilon(1S)$ at $\sqrt{s_{NN}} = 2.76$ TeV.  (Right) Same at $\sqrt{s_{NN}} = 5.02$ TeV.  
}
\label{fig:3}
\end{figure}

\begin{figure}[t!]
\centerline{
\includegraphics[width=0.485\linewidth]{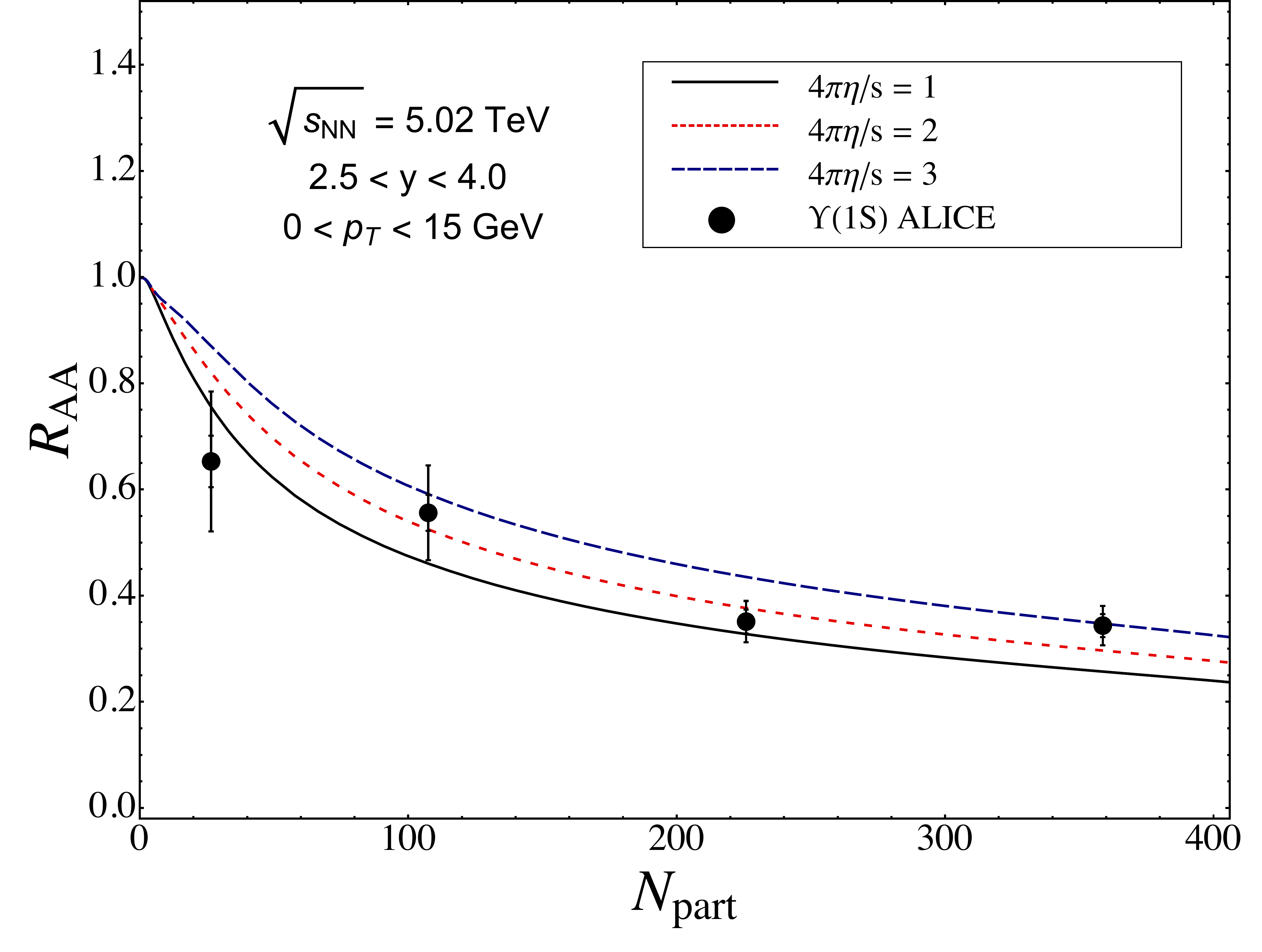}\hspace{2mm}
\includegraphics[width=0.465\linewidth]{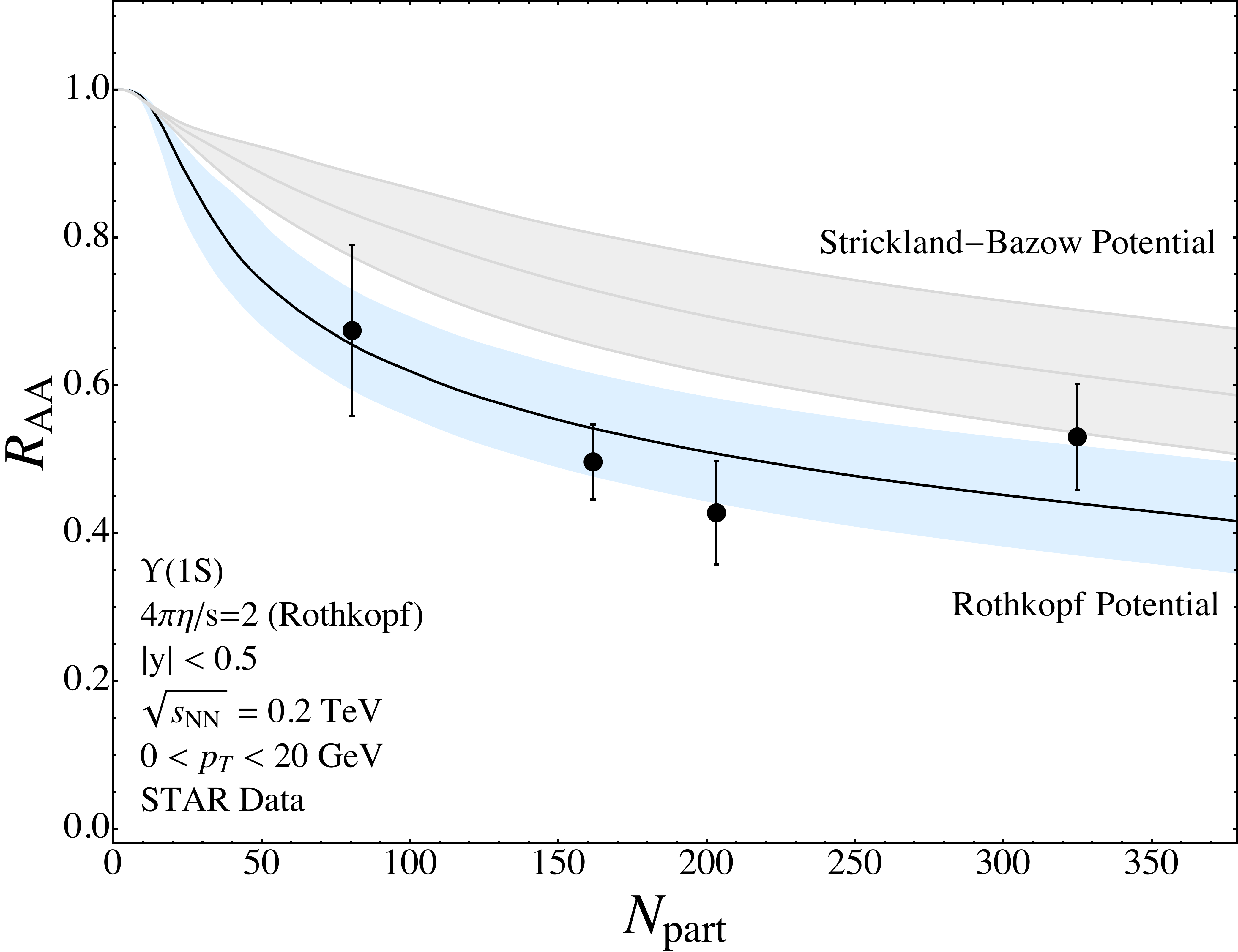}
}
\caption{(Left) $R_{AA}$ versus $N_{\rm part}$ for $\Upsilon(1S)$ at $\sqrt{s_{NN}} = 5.02$ TeV compared to preliminary ALICE data \cite{ALICE5TeV}.  (Right) $R_{AA}$ versus $N_{\rm part}$ for $\Upsilon(1S)$ at $\sqrt{s_{NN}} = 200$ GeV compared to preliminary STAR data \cite{STARupsilon}.
}
\label{fig:4}
\end{figure}

In Fig.~\ref{fig:3} we show the rapidity dependence of the $\Upsilon(1s)$, $\Upsilon(2s)$, and $\Upsilon(3s)$ $R_{AA}$.  The left plot shows model predictions for 2.76 TeV collisions together with combined CMS and ALICE data.  The right plot shows the same at 5.02 TeV.  Note that in both plots one sees a small increase in the KSUa model $R_{AA}$ at central rapidity which is associated with the effect of regeneration.  In the left plot we see that the KSUa model does not agree with the ALICE $\Upsilon(1s)$ data points (open circles), however, the model does a reasonable job describing both the CMS and ALICE rapidity dependence at 5.02 TeV (right) and the rapidity dependence of the CMS data at 2.76 TeV (left).

In Fig.~\ref{fig:4} we show results from two different collision energies.  The left plot shows the $N_{\rm part}$ dependence of the $\Upsilon(1s)$ $R_{AA}$ for 5.02 TeV collisions compared to data reported by the ALICE collaboration \cite{ALICE5TeV}.  Once again we see that the KSUa model does a quite good job description the centrality dependence of the observed $\Upsilon(1s)$ suppression.  Finally, turning to the right plot we compare the KSUa model and the KSUa model using the Rothkopf potential \cite{Krouppa:2017jlg,Krouppa:2018lkt} with preliminary experimental data collected by the STAR collaboration \cite{STARupsilon}.\footnote{The Rothkopf model uses the same background evolution and adiabatic approximation as the KSUa model, but uses a different underlying complex potential which is constrained in equilibrium by lattice QCD calculations~\cite{Krouppa:2017jlg}.}. Compared to the preliminary data it seems that the KSUa model under-predicts the amount of suppression observed by STAR, however, the Rothkopf potential does a better job.  This, however, comes at a cost since the Rothkopf potential overpredicts the amount of suppression at higher collision energies \cite{Krouppa:2017jlg,Krouppa:2018lkt}.


\subsection{First steps beyond the adiabatic approximation}

In order to go beyond the adiabatic approximation one can instead solve the time-dependent Schr\"odinger equation with a time-dependent complex in-medium heavy quarkonium potential.  In a recent paper \cite{Boyd:2019arx}, it was demonstrated that one can numerically solve the Schr\"odinger equation in real-time using an efficient split-operator method and extract the survival probability directly from the final quantum wave function.  In this case, one uses time-dependent eigenstates and computes the overlap of these with the time-evolved wave function.  

For a radial potential, the starting point is the expansion of the quantum mechanical wavefunction~\cite{Boyd:2019arx}
\begin{equation}
u(r,\theta,\phi,t) = {\cal N} \sum_{\ell=\ell_{\rm min}}^{\ell_{\rm max}} \sum_{m=-\ell}^\ell \frac{1}{\sqrt{2\ell+1}} u_\ell(r,t) Y_{\ell m}(\theta,\phi) \, ,
\end{equation}
from which one derive the following evolution equation
\begin{equation}
u_\ell(r,t + \Delta t) = \exp(- i \hat{H}_\ell \Delta t) u_\ell(r,t) \, ,
\label{eq:uUpdate}
\end{equation}
with $\hat{H}_\ell = \frac{\hat{p}^2}{2m} + V_{{\rm eff},\ell}(r,t)$ and $V_{{\rm eff},\ell}(r,t) = V(r,t) + \frac{\ell(\ell+1)}{2 m r^2}$ and $\hat{p} = -i \frac{d}{dr}$.

Once cast into this form, one can use the approximate time evolution operator which follows from application of the Baker-Campbell-Hausdorf formula 
\begin{equation}
\exp(-i \hat{H_\ell}(t) \Delta t) \simeq \exp(- i V_{{\rm eff},\ell}(r,t) \Delta t/2) \exp\!\left(-i \frac{\hat{p}^2}{2 m} \Delta t\right) \exp(- i V_{{\rm eff},\ell}(r,t) \Delta t/2) + {\cal O}( (\Delta t)^2 ) \, . 
\end{equation}

Since $u_\ell(r,t)$ must vanish at the origin, one can use a real-valued Fourier sine series to describe both the real and imaginary parts of the wave function.  This guarantees that the correct boundary conditions at $r=0$ are satisfied automatically and allows one to easily exclude the point $r=0$.  The resulting update steps are 
\begin{enumerate}
\item Update in configuration space using a half-step: $ \psi_1 = \exp(- i V_{{\rm eff},\ell} \Delta t/2) \psi_0$.
\item Perform Fourier sine transformations on real and imaginary parts separately:  \\ $\tilde\psi_1 = \mathbb{F}_s[\Re \psi_1] + i \mathbb{F}_s[\Im \psi_1] $.
\item Update in momentum space using: $\tilde\psi_2 =  \exp\!\left(-i \frac{p^2}{2 m} \Delta t\right) \tilde\psi_1$.
\item Perform inverse Fourier sine transformations on real and imaginary parts separately:  \\ $\psi_2 = \mathbb{F}_s^{-1}[\Re \tilde\psi_2] + i \mathbb{F}_s^{-1}[\Im \tilde\psi_2] $.
\item Update in configuration space using a half-step: $ \psi_3 = \exp(- i V_{{\rm eff},\ell} \Delta t/2) \psi_2$.
\end{enumerate}
After these steps are complete $\psi_3$ contains the updated wave function.  One then proceeds by iterating this procedure.  Note that in the case that the potential is real-valued this method exactly preserves unitarity of the quantum wave function.  In general, if the potential is time dependent, then there will be quantum state mixing since the instantaneous wave functions change in time.  This leads to the possibility of quantum production of excited states as we will see below.

One can initialize the system with a linear superposition of self-consistently determined in-medium bottomonium wave functions and then evolve the wave function to a given time.  At any time during the evolution one can project on the instantaneous wave functions of the time-dependent potential.  These are determined by a modified point-and-shoot method~\cite{Boyd:2019arx}.  Having obtained the full numerical solution, one can compare the resulting survival probabilities with those obtained using the adiabatic approximation with the same underlying complex potential.

\subsubsection*{Results}

\begin{figure}[t!]
\centerline{
\includegraphics[width=0.47\linewidth]{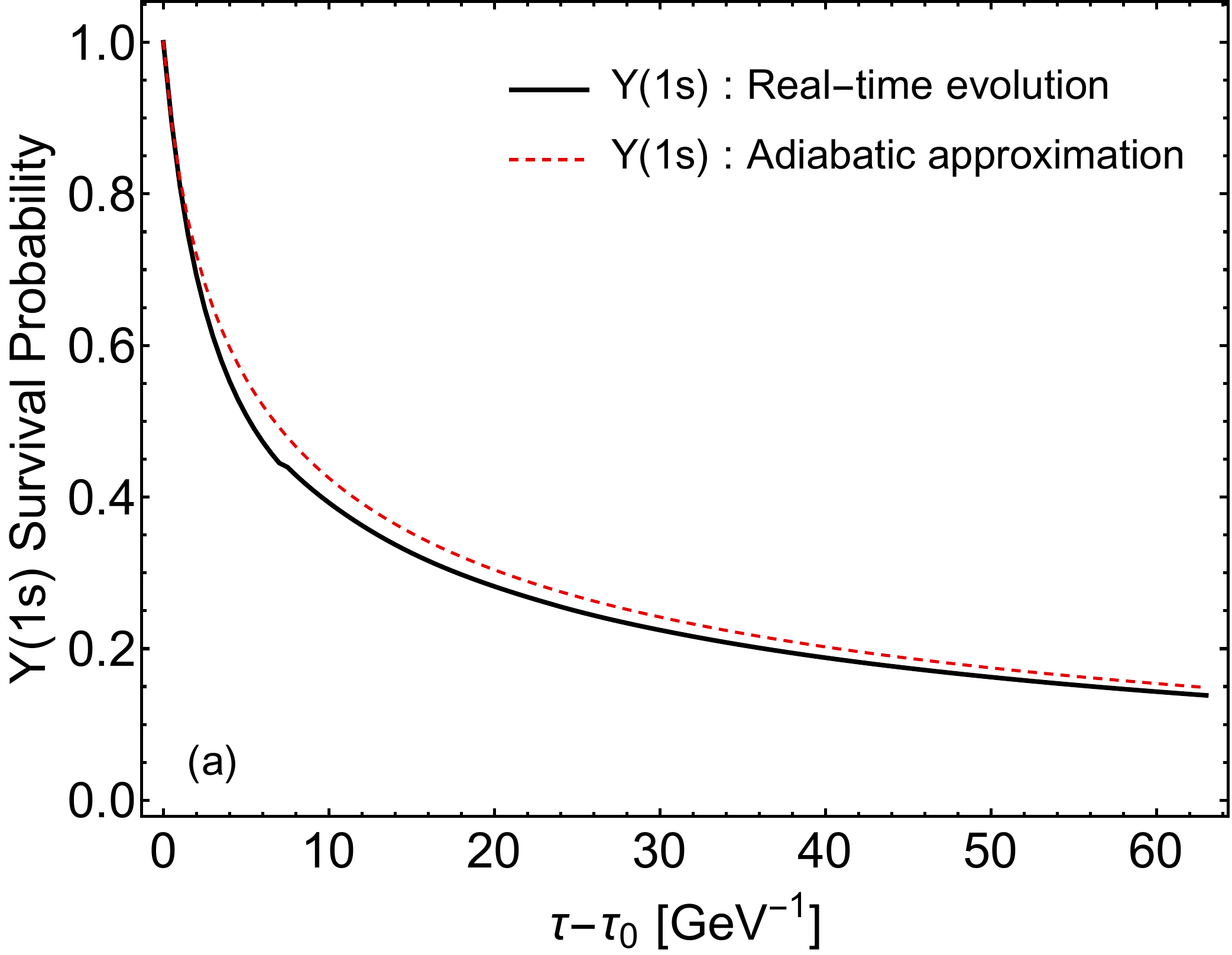}\hspace{2mm}
\includegraphics[width=0.495\linewidth]{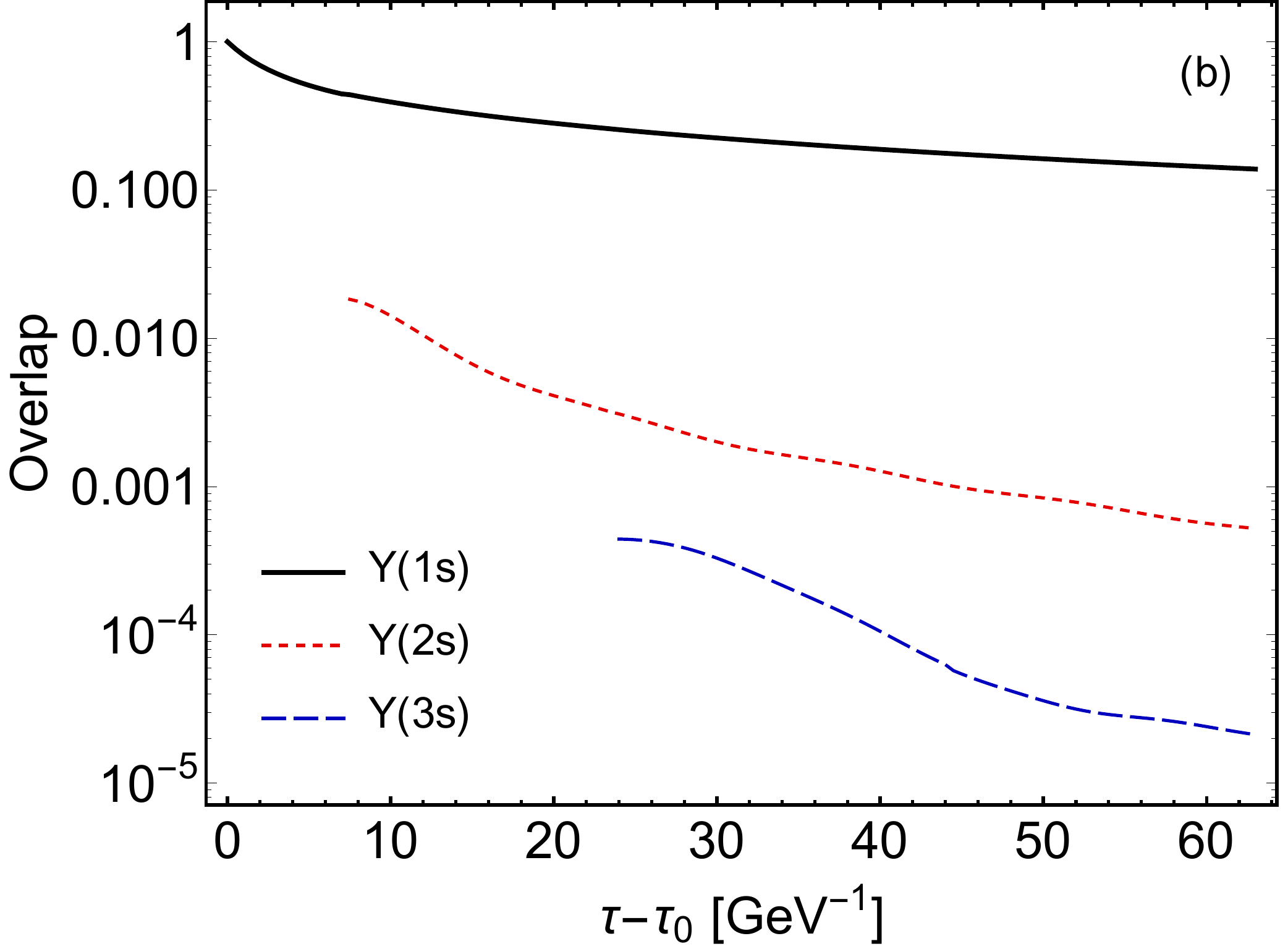}
}
\caption{Left plot shows the survival probability of the $\Upsilon(1s)$ as a function of proper time.  The right plot shows the overlaps extracted using the real-time evolution.
}
\label{fig:y1s0p6}
\end{figure}

For the background, Ref.~\cite{Boyd:2019arx} considered a transversally homogeneous and boost-invariant ideal fluid with a temperature which decreases as $T(\tau) = T_0 (\tau_0/\tau)^{1/3}$.  In Fig.~\ref{fig:y1s0p6} we present a comparison of the $\Upsilon(1s)$ survival probability obtained using the adiabatic model (KSUa) and the real-time evolution (KSUd).  The left plot shows the survival probability of the $\Upsilon(1s)$ as a function of proper time.  In this panel, the solid black line is the result obtained using real-time evolution and the dashed red line is the result obtained using the adiabatic approximation.  The right panel shows the overlaps extracted using the real-time evolution.  In this panel, the solid black, shorted-dashed red, and long-dashed blue lines are the overlaps computed for the 1s, 2s, and 3s states.  In both panels the initial temperature was taken to be $T_0 = 0.6$ GeV at  $\tau_0 = $ 1 GeV$^{-1}$ and the initial wave function consisted of a 1s state with the wave function determined self-consistently using the complex in-medium heavy quarkonium potential.

As we can see from the left panel of Fig.~\ref{fig:y1s0p6} the correction to the adiabatic evolution is small in this case, which corresponds physically to the evolution in the center of the QGP created in LHC 2.76 TeV collisions.  From the right panel, however, we see that, due to the time dependence of the in-medium potential, it is possible to dynamically generate overlaps with the excited states.  In this case, we see overlaps with both the $\Upsilon(2s)$ and $\Upsilon(3s)$ appearing, but they are quite small and would represent only a small correction to $\Upsilon(2s)$ and $\Upsilon(3s)$ production.

\begin{figure}[t!]
\centerline{
\includegraphics[width=0.485\linewidth]{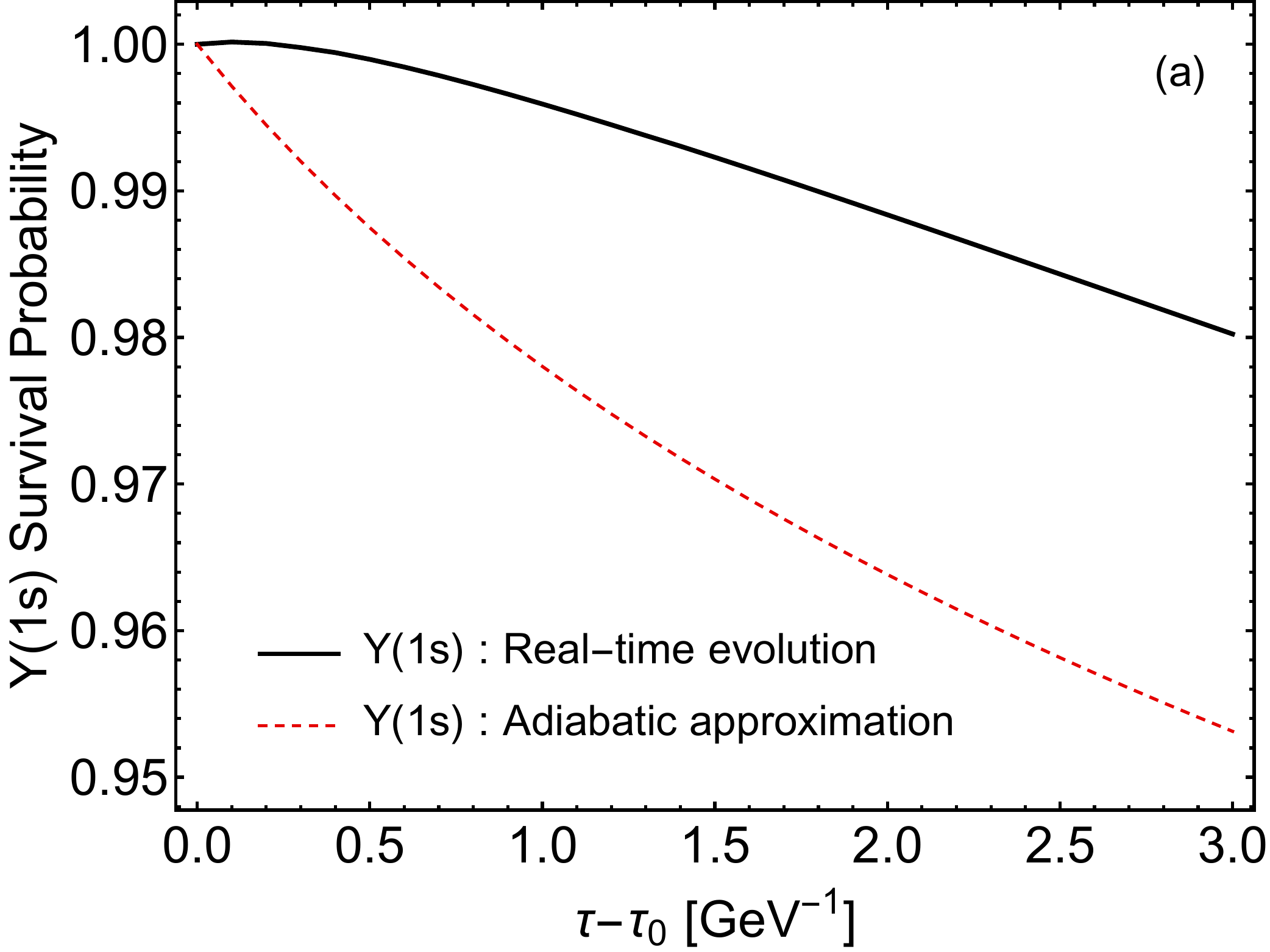}
\includegraphics[width=0.465\linewidth]{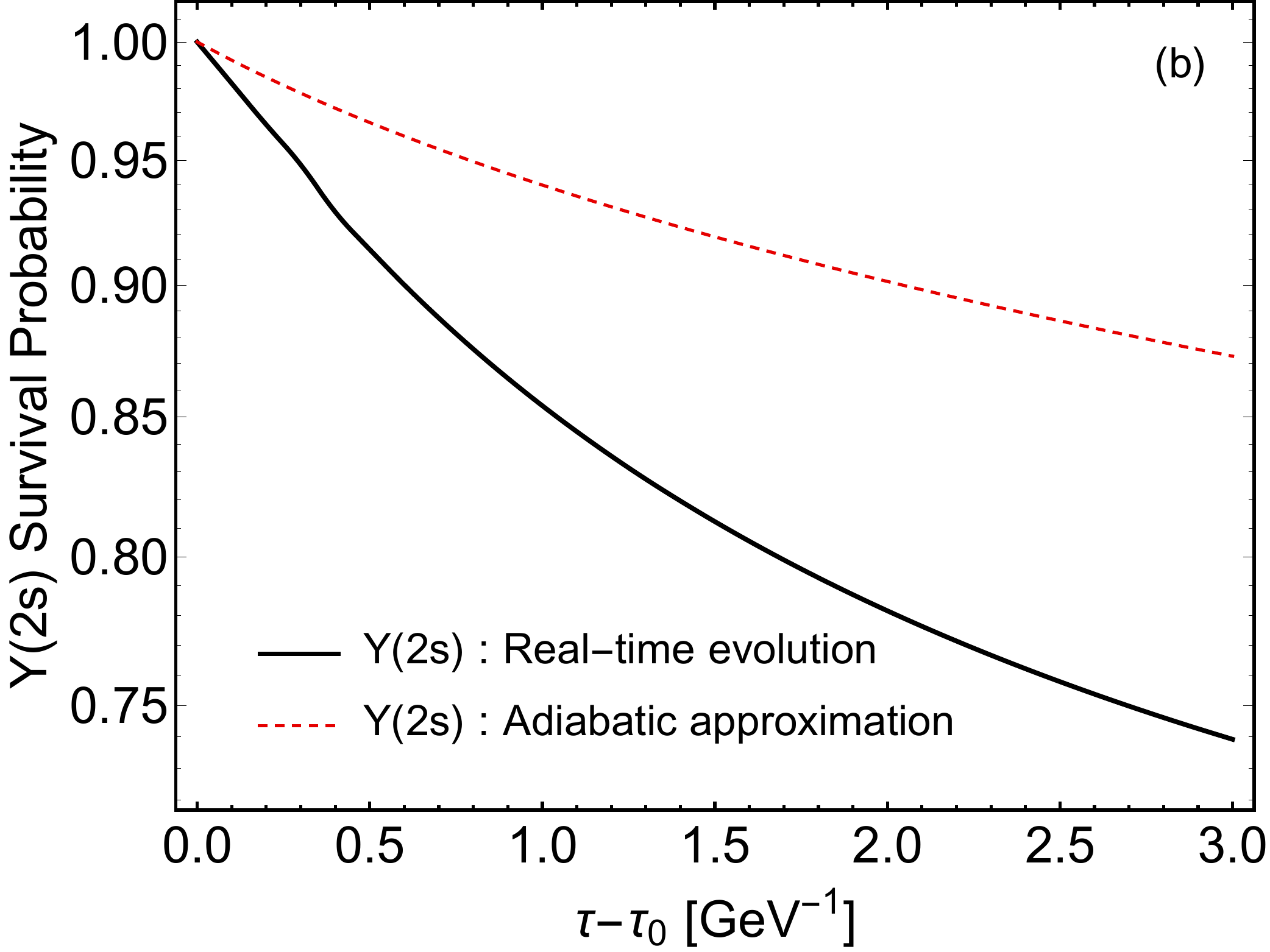}
}
\caption{(Left) $R_{AA}$ versus $N_{\rm part}$ for $\Upsilon(1S)$ at $\sqrt{s_{NN}} = 5.02$ TeV compared to preliminary ALICE data \cite{ALICE5TeV}.  (Right) $R_{AA}$ versus $N_{\rm part}$ for $\Upsilon(1S)$ at $\sqrt{s_{NN}} = 200$ GeV compared to preliminary STAR data \cite{STARupsilon}.
}
\label{fig:y1s0p225}
\end{figure}

In Fig.~\ref{fig:y1s0p225} we present similar comparisons for a lower initial temperature consistent with a peripheral heavy-ion collision or the periphery of a central heavy ion collision at RHIC or LHC, $T_0 = 0.225$ GeV at  $\tau_0 = $ 1 GeV$^{-1}$.  At this temperature, both the $\Upsilon(1s)$ and $\Upsilon(2s)$ are bound.  As a result, the quantum state of the system was initialized as a linear superposition with the relative probability set by the ratio of their production cross sections in $pp$ collisions.  The left plot shows the $\Upsilon(1s)$ survival probability and the right plot shows the $\Upsilon(2s)$ survival probability.  For the $\Upsilon(1s)$ there are visible corrections to the adiabatic evolution, however, they are still quantitatively on the order of a few percent.  In contrast, the right plot shows that the $\Upsilon(2s)$ correction can be as large as 18\%, with the real-time evolution giving enhanced suppression. 

\section{Conclusions}

In this proceedings contribution I presented a brief review of the progress that has been made in recent years to understand the suppression of bottomonium bound states in heavy-ion collisions.  In the first part,  I presented the "standard" KSUa model which builds upon advances in the calculation of in-medium heavy quarkonium evolution using finite temperature quantum field theory and effective field theory.  These first principles calculations are then folded together with a realistic 3+1d anisotropic hydrodynamics background which has been independently tuned to soft observables \cite{Alqahtani:2017jwl,Alqahtani:2017tnq,Almaalol:2018gjh}.  I demonstrated that the KSUa model provides a quite good description of the data collected by both ALICE and CMS at LHC, however, it seems to under-predict the amount of $\Upsilon(1s)$ suppression at RHIC energies based on comparisons to preliminary STAR data.  In the second part, I presented some recent work on relaxing the adiabatic approximation.  The results shown were for the academic case of Bjorken expansion, but the comparisons allowed us to gauge the order of magnitude of the error made when taking the adiabatic limit.  We are currently working on extending the treatment to 3+1d in order to determine the final phenomenological impact.

\section*{Acknowledgments} 

\noindent
I thank my collaborators and students for their help in performing the research reported herein.  This work was supported by the U.S. Department of Energy, Office of Science, Office of Nuclear Physics under Award No. DE-SC0013470.

\bibliographystyle{utphys}
\bibliography{strickland}

\end{document}